\documentclass[floatfix,prd,lengthcheck,showpacs,amssymb,amsmath,amsfonts,aps,altaffilletter,nofootinbib,nopreprintnumbers,showpacsm]{revtex4-1}

\usepackage[utf8]{inputenc}
\usepackage[dvipsnames]{xcolor}
\usepackage{graphicx}
\usepackage{subfigure}
\usepackage{amsmath}
\usepackage{acronym}
\usepackage{multirow}
\usepackage{tabu}
\usepackage{import}
\usepackage{hyperref}
\usepackage[normalem]{ulem}
\usepackage{soul}
\usepackage{calrsfs}
\usepackage{caption}
\usepackage{lineno}
\DeclareMathAlphabet{\pazocal}{OMS}{zplm}{m}{n}

\hypersetup{
    colorlinks=true,
    linkcolor=blue,
    filecolor=magenta,      
    urlcolor=cyan,
}

\begin{document}

\title[]{Unveiling gravitational waves from core-collapse supernovae with MUSE}

\author{A. Veutro$^{1a,b}$, I. Di Palma$^{1a,b}$, M. Drago$^{1a,b}$, P. Cerd\'{a}-Dur\'{a}n$^2$, R. van der Laag $^3$,  M. L\'{o}pez $^{4,5}$, F. Ricci$^{1a,b}$}

\affiliation{$^{1a}$ Universit\`a di Roma  {\it{La Sapienza}}, I-00185 Roma, Italy}
\affiliation{$^{1b}$ INFN, Sezione di Roma, I-00185 Roma, Italy}
\affiliation{$^2$ Departamento de Astronom\'ia y Astrof\'isica, Universitat de Val\`encia, Dr. Moliner 50, 46100, Burjassot (Valencia), Spain}
\affiliation{$^3$ Leiden Institute of Advanced Computer Science, Leiden University, Einsteinweg 55, 2333CC Leiden, The Netherlands}
\affiliation{$^4$ Institute for Gravitational and Subatomic Physics (GRASP), 
Department of Physics, Utrecht University, 
Princetonplein 1, 3584 CC Utrecht, The Netherlands}
\affiliation{$^{5}$ Nikhef, Science Park 105, 1098 XG Amsterdam, The Netherlands}



\begin{abstract}
The core collapse of a massive star at the end of its life can give rise to one of the most powerful phenomena in the Universe. Because of violent mass motions that take place during the explosion, core-collapse supernovae have been considered a potential source of detectable gravitational waveforms for decades. However, their intrinsic stochasticity makes ineffective the use of modelled techniques such as matched filtering, forcing us to develop model independent technique to unveil their nature. In this work we present MUSE pipeline, which is based on a classification procedure of the time-frequency images using a Convolutional Neural Network. The network is trained on phenomenological waveforms that are built to mimic the main common features observed in numerical simulation.
The method is finally tested on a representative 3D simulation catalog in the context of Einstein Telescope, a third generation GW telescope. Among the three detector geometries considered here, the 2L with a relative inclination of $45^\circ$ is the one achieving the best results, thus being able to detect a Kuroda2016-like waveform with an efficiency above $90\%$ at 50 kpc.
\end{abstract}

\maketitle

\section{Introduction}
\label{sec:introduction}

The first detection of a binary black hole {(BBH)} merger in 2015 {by LIGO \cite{LIGOScientific:2014pky}  and Virgo \cite{ VIRGO:2014yos} collaborations} have marked the birth of the gravitational wave (GW) astronomy \cite{PhysRevLett.116.061102}. Since then, more than one hundred detections have been confirmed so far and almost all of them have been associated to BBH merger \cite{LIGOScientific:2018mvr, LIGOScientific:2021usb, KAGRA:2021vkt}. Among the few exceptions, the most remarkable is GW170817, a binary neutron star (BNS) merger, that was observed in coincidence with the electromagnetic detection of a short gamma-ray burst, thus opening a new era of multi-messenger astronomy, in which GWs played a crucial role \cite{Goldstein_2017, Abbott_2017, Yang_2019}.

The continuous improvements of currently operating GW detectors, together with the planned next generation facilities (Einstein Telectope \cite{ET-0007C-20}, Cosmic Explorer \cite{Reitze:2019iox}) will led not only to an increase of the number of detections, but also to the observation of novel sources of GWs. Among them, one of the most interesting is undoubtedly core-collapse supernova (CCSN): a powerful explosion triggered by the collapse of massive stars, which is expected to generate GWs due to aspherical mass motions inside the source, even if the majority of its energy is released via neutrinos, as it was confirmed through the detection of MeV neutrinos from SN1987A \cite{Li:2023ulf}.

The collapse of massive stars ($M>8-10M_\odot$) is initiated when the star exhausts its nuclear fuel leaving behind a central iron core. With fusion no longer able to counteract gravity, the core of the
star collapses inwards within milliseconds until the nuclear saturation density is reached ($\rho = 2.7 \, \times \, 10^{14} $g \, cm$^{-3}$). The repulsive contributions of the nuclear force between the nucleons, halts the collapse of the inner core forming a proto-neutron star (PNS). The remaining infalling matter bounces back and its expansion creates pressure waves that steepen into a shock front, which stalls at about 100 km from the core after having lost its energy via photo-dissociation of iron nuclei and neutrino cooling. 
Accordingly to the neutrino-driven mechanism \cite{Bethe:1990mw}, the energy deposited by neutrinos emitted by the PNS can help revive the shock and allow the explosion to set in, producing the so-called supernova {(SN)}. This paradigms successfully explains explosions of slowly rotating progenitors. On the other hand, SN explosion of rapidly rotating progenitors is well described by the magneto-rotational mechanism, in which strong magnetic fields play a crucial role in pushing the SN shock and producing highly energetic events, like hypernovae and long GRBs \cite{1992ApJ...392L...9D, 1976Ap&SS..41..287B, Akiyama_2003, nature1992}. Because only around 1 \% of the events shows signatures of fast rotation, in this work we will focus only on the slowly-rotating SNe.

In a {SN} explosion, GWs are generated in the inner core of the source, thus carrying direct information about the inner mechanism of this phenomen. However, {although} they are one of the most energetic events in the universe, the expected GW emission is too weak to be detected at extra-galactic distances, i.e. the expected GW strain amplitude from a CCSN in the center of Milky Way ranges between $10^{-21}$ and $10^{-23}$. Furthermore, due to the high-level of complexity, the predicted GW signals are intrinsically stochastic, making ineffective the use of {modelled techniques such as matched filtering} \cite{Messick:2016aqy, hanna2020fast, Sachdev:2019vvd, usman2016pycbc, dal2014implementing, Allen:2005fk, nitz2017detecting}. 
An attempt of building a matched filter template, from SN neutrino event rate, for detecting SASI modulations in GW signal has been made, however no significant improvements in the overall GW detection efficiency have been found \cite{Drago:2023cve}.
The current approaches used in CCSNe searches employ model-free algorithms that rely on excess of power to identify signals buried in detector noise, such as coherent Waveburst (\texttt{cWB}) \cite{Drago:2020kic}, but these methods do not leverage of any specific feature of CCSN waveform.

Machine learning {(ML)} techniques can tackle these issues thanks to their adaptability and rapidity since all the intensive computation is diverted to the one-time training stage, which could make them orders of magnitude faster than conventional matched filtering technique \cite{Gabbard:2017lja}.
In recent years, the GW community has actively investigated various ML applications in data analysis, including compact binary coalescence (CBC) detection \cite{George:2016hay, PhysRevD.105.043006, ruder2016overview, Gabbard:2017lja, Sharma:2022ibm, Krastev, Menendez-Vazquez:2020khz, Baltus:2021nme, Baltus:2022pep, 2024arXiv240905068M, Nousi:2022dwh, Koloniari:2024kww}, burst event identification \cite{2020arXiv200204591C, PhysRevD.105.084054, Iess:2020yqj, Iess:2023quq, Lopez:2021rci, Meijer:2023yhn, Boudart:2022xib, Boudart:2022apz, Skliris:2020qax}, glitch classification \cite{Biswas:2013wfa, zevin2017gravity, Glanzer:2022avx, bahaadini2018machine, Laguarta:2023evo}, and synthetic data generation \cite{Lopez:2022lkd, Lopez:2022dho, Powell:2022pcg,Yan:2022spw, Dooney:2022arh, Dooney:2024pvt}, among other areas. For a comprehensive overview, we direct interested readers to the reviews in \cite{Cuoco:2020ogp} and \cite{Schafer:2022dxv} \newline

In the context of CCSN search, \cite{Astone:2018uge} and \cite{LopezPortilla:2020odz} {introduce} a novel and promising method to detect GWs signal from CCSNe using ML techniques. 
In \cite{Astone:2018uge}, they show that the model achieves slightly better performance than \texttt{cWB}, while in \cite{LopezPortilla:2020odz} they apply the method to LIGO Livingston, LIGO Hanford and Virgo real data collected during the second observation run (O2), reaching an efficiency of detection around 60\% at 15 kpc.
In this paper, the Multi-messenger Understanding of Supernova Explosions (MUSE) pipeline is presented. 
MUSE adopts the same deep neural network architecture developed in \cite{LopezPortilla:2020odz}. However, instead of using the original \texttt{TensorFlow}/\texttt{KERAS} framework, it uses \texttt{PyTorch}, which offers several advantages including a simpler and more flexible network customization and a more granular control over the training process. Moreover, the ML dataset preparation has been updated through PyCBC \cite{alex_nitz_2024_10473621}. Both these changes have lead to a significant decrease of the total computational time, as discussed in Section \ref{sec:muse}. Here, the performances of the MUSE pipeline are shown for a representative dataset of 3D neutrino-driven (non-rotating) CCSN simulations in the context of Einstein Telescope (ET), a third generation GW detector.\newline

The paper is structured as it follows. In Section \ref{sec:convolutional}  we provide a brief introduction to Deep Neural Networks with a focus on Convolutional Neural Networks. In Section \ref{sec:phenomenological} we describe the code used to generate phenomenological waveforms that can mimic the peculiarities of the CCSN GW signal observed in numerical simulations. In Section \ref{sec:muse} there is a detailed description of the MUSE pipeline. In Section \ref{sec:dataset} we describe each dataset. In Section \ref{sec:results} we show, as a study case, the result of testing MUSE on ET using a representative set of waveforms taken from 3D CCSN simulations.

\section{Convolutional neural networks}
\label{sec:convolutional}

Deep learning {(DL)} is a subset of {ML} that has revolutionized the field of artificial intelligence by enabling computers to learn and make decisions with unprecedented accuracy and complexity. Inspired by the human brain's neural networks, {DL} employs Deep Neural Networks (DNNs)--a multi-layered Artificial Neural Networks (ANNs)--to model intricate patterns and relationships within the data \cite{483329}. 

By definition, an ANN is an  approximation function that maps inputs $\textbf{x} \in \mathbb{R}^n$ to outputs $\textbf{y} \in \mathbb{R}^m$ through layers of neurons, using linear transformations followed by non-linear activation functions. Given an input vector $\textbf{x}$, the output $a_i$ of the $i^{th}$ neuron is then:
\begin{equation}
    z_i = \sum\limits_j w_{ij}x_j + b_i
\end{equation}
\begin{equation}
    a_i = \phi(z_i)
\end{equation}
where $\phi$ is the activation function ($e.g.$ ReLU, sigmoid \cite{2022arXiv220402921G}), $b_i$ is the bias term and $w_{ij}$ are the weights. $b_i$ and $w_{ij}$ are trainable parameters, which means that they are not fixed but adjusted dynamically by minimizing the loss function: a mathematical formula that measures how well a neural network's predictions match the actual target values through iterative optimization, typically using gradient-based methods \cite{2016arXiv160904747R}. \newline
Aside from these learnable parameters, there is set of parameters not learnt during the training but fixed at the beginning of it, so-called hyper-parameters. Some common hyper-parameters are,  
\begin{itemize}
\item \textit{Network architecture:} a neural network is composed of multiple layers of neurons organized into three main types: 
\begin{enumerate} 
    \item  the input layer,
    \item one (ANN) or more (DNN) hidden layers, whose representations are not directly observed,
    \item the output layer, which produces the final prediction or classification.
\end{enumerate}
Depending on the way this layers are characterized, various DNN architectures can be defined.
\item \textit{Learning rate ($lr$):}  controls the step size at which a neural network updates its weights during the training phase using gradient descent.
\item \textit{Batch size:} defines the number of training samples processed simultaneously before the model’s internal parameters are updated.
\item \textit{Positive class weight $w_{pos}$}: used to assign higher ($w_{pos} > 1$) or lower ($w_{pos} < 1$) loss weight to the positive class, so that the model pays more attention to positive or negative class, respectively.
\item \textit{Optimizer}: gradient descent strategy to minimise the cost function of the the model.
\end{itemize}
Among the different DNN architectures, convolutional neural network (CNN) is a class of DL architectures designed to process data with grid-like topology, such as images or time-series signals. CNNs are particularly effective in tasks that require automatic feature extraction, pattern recognition, and classification, due to their ability to learn spatial and temporal hierarchies through layers of convolutional filters. Convolutional filters, also known as kernels, are small, learnable matrices that are applied across the input data to extract local features present in the image. These local features are identified in a hierarchical way: early layers typically capture simple structures, while deeper layers learn increasingly abstract and complex representations of the input. During training, filters are adjusted by the network to detect features that are most informative for the given task. After each convolution filter, a pooling layer could be used. The aim is to down-sample the output from the convolutional layer, simplifying the information and reducing computational cost by decreasing the number of learnable parameters. This makes the model symmetric for pattern translation inside the input image, since once a sub-feature is found knowing the exact position is not as important as knowing the relative position with respect to the other sub-features in the image. 
Following the convolution blocks, the output of the convolutional layers is connected via a flattening layer with one or more dense layers (DNN). In cases of classification task, this approach is used to extract a score or probability, which is a numerical measure that reflects how well a classification model is performing in assigning input data to the correct classes. \newline 

Because of the effectiveness of ML algorithms in identifying patterns in data, they can be exploited to make GW searches more sensitive and robust. This is  especially interesting in the context of unmodelled GW searches, where the target sources are generally unmodelled due to either unknown theoretical background and/or complex dynamics of the system. ML techniques do not have inherent constraints on the size of the template bank to be used for training the network; in fact, using large datasets is advantageous, as it allows for broader coverage of the parameter space and enhances the model's ability to generalize across a wide range of signal morphologies. However, at present, multidimensional numerical simulations of CCSN waveforms yield only a limited set due to the substantial computational cost. At the same time, the progenitor models employed in these simulations may introduce biases, as many are specifically designed to reproduce features observed in supernova SN1987A. Since machine learning requires a large amount of data to develop generalization capabilities, we use parameterized phenomenological waveforms that resemble real gravitational waves - specifically, those derived from numerical simulations — and evaluate the model’s generalization performance on a blind dataset of numerical simulation outcomes.

\section{Phenomenological Waveforms}
\label{sec:phenomenological}
To train our network, we use labeled data samples that consist of either detector noise alone or detector noise with an added CCSN signal, referred to as injections. A number of works in the literature have modeled the GW signature of CCSN through three-dimensional numerical hydrodynamics simulations \cite{Kuroda:2016, Andresen:2017,Kuroda2017,Yakunin2017,O'Connor2018,Andresen2019,Powell2019,Radice2019,Powell2020,Pan2021,Powell2021,Mezzacappa2020,Mezzacappa2023,Vartanyan2023}. These simulations include some neutrino transport treatment, general relativistic corrections to the gravitational field and consider a wide range of progenitor stars with varying mass, metallicity and rotation rate. In this work we consider non-rotating progenitors, which are expected to represent the vast majority of CCSN events (only $1\%$ of the events show signatures of fast rotation such as broad-lined type Ic SNe \cite{Li2011} or long GRBs \cite{Chapman2007}).

Due to the high computational cost, the total number of three-dimensional simulations available in the literature is of the order of hundreds. This number is clearly too low for DNN training, validation and testing. In \cite{Astone:2018uge} and \cite{LopezPortilla:2020odz} the authors proposed the use of phenomenological waveforms for the training process. These waveforms are computationally inexpensive as a single waveform takes $\sim 10$ ms, enabling the fast generation of datasets of $\sim 100000$ waveforms for training and validation. Networks trained on phenomenological waveforms, detected numerical simulation signals during testing without much performance degradation \cite{LopezPortilla:2020odz}. Recently \cite{Cerda-Duran2025} have developed an improved version of the phenomenological waveform generator (ccphen v4) that will be used in this work.

The phenomenological waveform generator ccphen v4\footnote{Plublicly available at \url{https://www.uv.es/cerdupa/codes/ccphen/}.} models the GW emission as set of damped harmonic oscillators with varying frequency and a randomly generated forcing term. This generates waveforms, including both polarizations, that mimic the random phase present in waveforms from numerical simulations, but at the same time has a frequency evolution that resembles the spectrograms of realistic signals, in particular the raising arch pattern dominating the GW emission (associated to the excitation of the quadrupole oscillation modes of the PNS), and the possibility of having a weaker component at about $100$~Hz related to the standing shock accretion instability (SASI).

The waveforms are parametrized by a set of internal parameters including the signal duration, the emission delay with respect to the bounce, the quality factor $Q$ of the damping term (determining the broadness of the signal) and shape of the frequency evolution (determined by three reference frequencies). For signals with both a dominant component and a SASI component, we have a set of internal parameters for each component, plus the relative amplitude between them. Note that the range of possible values for each of the parameters, as well as the calibration of the waveform amplitude with the distance, has been determined by studying a set of three-dimensional numerical simulations \cite{Kuroda:2016, Andresen:2017,Kuroda2017,Yakunin2017,O'Connor2018,Andresen2019,Powell2019,Radice2019,Powell2020,Pan2021,Powell2021,Mezzacappa2020}. The calibration includes corrections due to the initial mass function of the progenitors to avoid biases coming from the selection of progenitors used in the simulations. External parameters of the waveform include the distance to source, its orientation in the sky (inclination and polarization angle) and its sky localization.
A detailed description of all the parameters, parameter ranges and calibration can be found in \cite{Cerda-Duran2025}.

ccphen v4 also provides an automatic waveform generator that allows to generate random waveforms within the calibrated parameter space for three different cases: signals including SASI, signals without SASI, and short signals (that mimic those from very low mass progenitors of $8-10 M_\odot$). For the scope of this work, only signals including SASI have been considered.

\begin{figure*}[htbp]
  	\centering

  	\subfigure[\,Kuroda et al. (2016) tapered waveform as outlined in Sec. \ref{sec:dataprep}.]{
  		\includegraphics[width=0.8\linewidth]{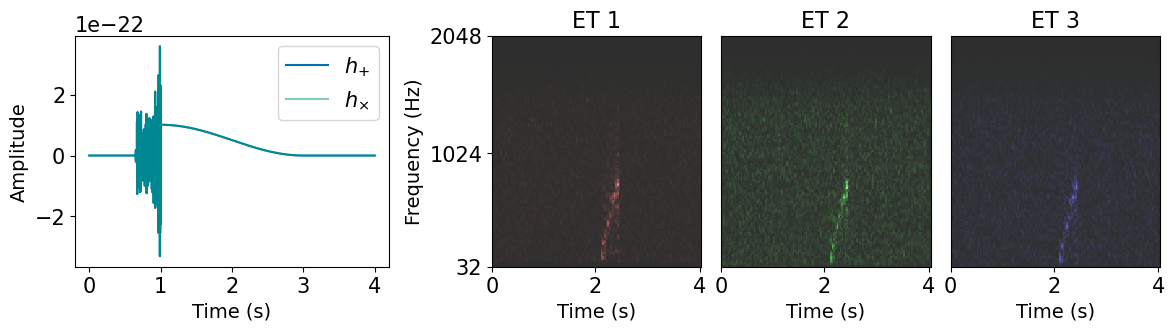}
		\label{fig:sub1}
  	}

	\vskip\baselineskip

  	\subfigure[\,ccphen v4 waveform with dominant mode and SASI.]{
    		\includegraphics[width=0.8\linewidth]{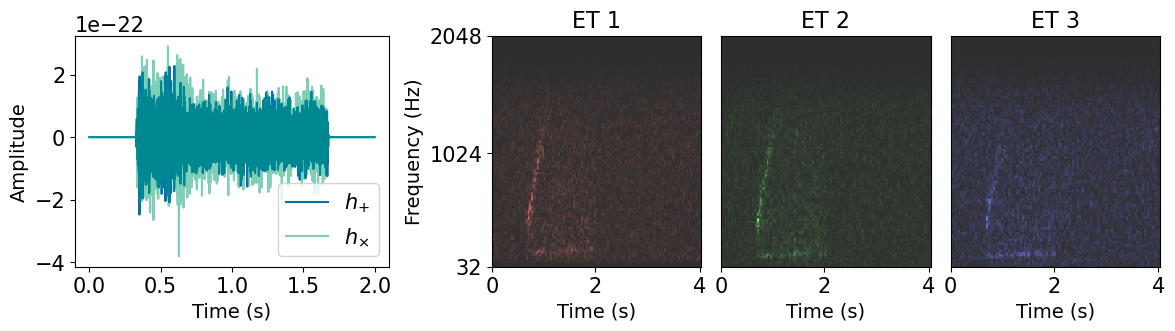}
    		\label{fig:sub2}
  	}

  	\caption{\textit{(Left)} Time series of $h_{+}$ and $h_{\times}$ representations of waveforms. \textit{(Right)} GW waveforms are embedded in design-sensitivity noise of the Einstein Telescope (ET), assuming a triangular configuration. We generate time-frequency spectrograms for each ET channel (ET1, ET2, ET3), using their respective RGB colors (red, green and blue) that will be the input to the neural network.}
  	\label{fig:combined}
\end{figure*}

\section{MUSE}
\label{sec:muse}

\subsection{Data preparation}
\label{sec:dataprep}
\begin{table*}[t]
    \centering
    \renewcommand{\arraystretch}{1.7}
    \begin{tabular}{ |w{c}{4cm}|w{c}{2cm}|w{c}{5cm}|w{c}{2cm}|  }
    \hline
    Authors & Year & ZAMS & $N_{sim}$ \\
    \hline
    Andresen et al. \cite{Andresen:2017} & 2016 & [11, 20, 27] M$_\odot$ & 4 \\
    Andresen et al. \cite{Andresen2019} & 2019 & 15 M$_\odot$ & 1 \\
    Kuroda et al. \cite{Kuroda:2016} & 2016 & 15 M$_\odot$ & 2 \\
    Kuroda et al. \cite{Kuroda2017} & 2017 & 15 M$_\odot$ & 2 \\
    Mezzacappa et al. \cite{Mezzacappa2020} & 2020 & 15 M$_\odot$ & 1 \\
    O'Connor \& Couch \cite{O'Connor2018} & 2018 & 20 M$_\odot$ & 6 \\
    Pan et al. \cite{Pan2021} & 2020 & 40 M$_\odot$ & 1 \\
    Powell et al. \cite{Powell2019} & 2018 & 18 M$_\odot$ & 1 \\
    Powell et al. \cite{Powell2020} & 2020 & [18, 20] M$_\odot$ & 2 \\
    Powell et al. \cite{Powell2021} & 2021 & [85, 100] M$_\odot$ & 5 \\
    Radice et al. \cite{Radice2019} & 2019 & [9, 10, 11, 12, 13, 19, 25, 60] M$_\odot$ & 8 \\
    Yakunin et al. \cite{Yakunin2017} & 2017 & 15 M$_\odot$ & 1 \\
    \hline
    \end{tabular}
    \caption{Summary of multi-dimensional simulations of non-rotating progenitor models employed in this study. \textit{Year} is the year of publication; \textit{ZAMS} is the Zero-Age Main-Sequence that refers to the mass of the progenitor; $N_{sim}$ is the amount of simulation outcomes extracted from each set.}
    \label{tab:1}
\end{table*}

In the following we provide a schematic description of the pre-processing of the input data before feeding it to the neural network.
\begin{enumerate}
    \item \textit{Waveform generation:}\newline
    Depending on the aim, signal waveforms can be generated using the phenomenological model described in Section \ref{sec:phenomenological} or they can be extracted from numerical simulation outcomes. The default sampling rate considered in MUSE is 4096 Hz. We resample numerical simulation data by first upsampling the strain through interpolation to a new time grid with a power of 2 number of points and then downsamplig it to the desired sampling frequency after applying a low pass filter at the Ninquist frequency\footnote{For this last step we use the function \texttt{scipy signal.decimate}, with a FIR filter of order $60$.}.

    \item \textit{Tapering:}\newline  
    CCSN GW waveforms from numerical simulations might have a non-zero at the end, either for physical or numerical reasons.  The abrupt jump from some finite value to zero will induce artifacts in the energy spectrum of the signal. Following \cite{2022PhRvD.105j3008R}, we can avoid  such artefacts by extending our signals with a function that tapers to zero over some finite time scale using Eq. 17. In Fig. \ref{fig:sub1} we can see the effect of tapering on a numerical relativity waveform in time series (left) and its time-frequency representation embedded in detector noise (right).
    
    \item \textit{Detector projection:}\newline
    Because we are dealing with laser interferometer detectors, which are not omnidirectional, but exhibit directional variations in sensitivity described by the ``antenna patterns", the signal waveform needs to be projected and weighted accordingly to the sky localization of the source. Knowing the two linear polarizations of a GW, plus ($h_+$) and cross ($h_\times$), and the antenna patterns $F_+$ and $F_\times$ of a given detector configuration, the resulting strain $h(t)$ measured at the interferometer is given by
    \begin{equation}
    h(t) = F_+(\theta,\phi)h_+ + F_\times(\theta,\phi)h_\times
    \label{projection}
    \end{equation}
    where $\theta$ and $\phi$ are the source coordinates in the sky. In MUSE {these source coordinates} are randomly sampled from uniform distribution. Moreover, the projected signal strain needs to be scaled accordingly to the source distance. In MUSE injection code there are three possible choices for the distance sampling distribution: (1) uniform, (2) exponential and (3) poissonian, each of them with their tunable free parameters.
    \item \textit{Injection in detector noise:}\newline
    The detector noise can be computed following two approaches: (1) ideal noise (Gaussian noise) generated with a frequency evolution according to the Power Spectral Density (PSD) of the detector or (2) real data taken directly from GWOSC\footnote{MUSE pipeline can input real data from O3. Refer to \url{https://gwosc.org/O3/o3_details/} for technical details.} \cite{RICHABBOTT2021100658}.
    \item \textit{Whitening procedure:}\newline
    The frequency dependence of detector PSD makes the search of power excess across a wide band for frequencies challenging. To this extent, it is crucial to whiten the data by normalizing the power at all frequencies and suppressing frequency-dependent noise components. In MUSE this is done using \texttt{PyCBC} library\footnote{In particular we use \texttt{pycbc.types.timeseries.TimeSeries.whiten} function.} \cite{alex_nitz_2024_10473621} taking a $50\,$s time window frame around the injection to compute the reconstructed PSD.
    \item \textit{Spectrogram generation:}\newline
    {Short-time Fourier transform {(STFT)}} is applied to the whitened data to generate a spectrogram that covers a frequency { $\in [32, 2048]\,$Hz } with a time duration of $4\,$s. Moreover, keeping in mind the aim of using these spectrograms to feed a neural network, it is crucial to normalize the brightness of each pixel, constraining it between 0 and 1. This is done by simply dividing the brightness of each pixel by the brightest one.
\end{enumerate}
The result is a set of $N$ images, one per each detector, composed of a grid of $129 \times 129$ pixels with brightness between 0 and 1. It is important to note that, to avoid biasing the network toward specific regions of the image, waveform injections are randomly shifted in time, with the constraint that the entire signal must be fully contained within the 4 second window. In terms of speed performance, this new method has gained a factor of about 7.5 with respect to the one used in \cite{LopezPortilla:2020odz}. The final step of data preparation creates a set of $N$ \texttt{h5} input files and fills them with an equal number of injections with similar network SNR, which is defined as the quadrature sum of each detector SNR. These files are then used to feed the network with progressively low network SNR samples, following the curriculum learning approach described in the next section.

\subsection{Mini Inception-Resnet}

MUSE is based on the DNN developed in \cite{LopezPortilla:2020odz} to perform a binary classification task: we want to distinguish between positive class (signal waveforms added in detector noise) and a negative class (pure noise). Because of that, a binary cross-entropy ($\pazocal{L}$) has been chosen as loss function.
\begin{equation}
    \pazocal{L} = -\frac{1}{N} \sum\limits_{i=1}^N [y_i \log(p_i) + (1-y_i)\log(1-p_i)]
    \label{loss}
\end{equation}
where $N$ is the number of observations, $y_i$ is the actual binary label (0 or 1) and $p_i$ is the predicted probability. With the aim of improving the performance in \cite{LopezPortilla:2020odz}, we have modified the architecture, facing two main obstacles:
\begin{enumerate}
\item \textit{Increase of computational cost:} the increase of network size implies the increase of the number of trainable parameters and, consequently, of the computational cost. 
\item \textit{Performance degradation:} the degradation problem leads to worse performances, due to the fact that information struggles to propagate through many layers when considering deeper networks.
\end{enumerate}
The first issue has been addressed by introducing an Inception network, which employ a sparsely connected architectures to reduce the large amount of parameters appearing after network enlargement. On the other hand, the degradation problem is faced by the use of Residual Networks (ResNets). The peculiarity of this network is that it uses skip connections, a feature to bypass intermediate layers by directly feeding inputs to deeper layers, facilitating gradient flow and enabling the effective training of very deep architectures. 

The resulting network is a reduced version of Inception-Resnet v1 \cite{inception} (Mini Inception-Resnet, hereafter) with a total number of trainable parameters of about 98k. Following \cite{LopezPortilla:2020odz}, the network training is performed adopting the curriculum learning approach: a training strategy in which a model is exposed to progressively more complex examples, allowing it to build foundational representations before tackling more difficult data, thereby improving convergence and generalization. In our case, the complexity of the dataset signal pattern is evaluated via the network signal-to-noise ratio (SNR), which reflects how prominently a signal stands out against instrumental and environmental noise in case of multiple detectors $k$.
\begin{equation}
    \text{netSNR} = \sqrt{ \sum\limits_k \int \frac{\tilde{h}^2}{S_k(f)}df}
    \label{netsnr}
\end{equation}
where $\tilde{h}$ is the projected waveform strain in the frequency domain and $S_k(f)$ the amplitude spectral density for detector $k$.
It follows that, during the training, the network is fed with progressively lower network SNR injections. \newline
The network has been implemented inside the MUSE code using \texttt{PyTorch} \cite{2019arXiv191201703P}, which offers several advantages over \texttt{TensorFlow}/\texttt{KERAS} \cite{tensorflow2015-whitepaper} used previously in \cite{LopezPortilla:2020odz}, including a dynamic computation graph that enables more intuitive and simpler model customization.
Apart from the different DL framework adopted with respect to \cite{LopezPortilla:2020odz}, the main differences with this previous work are:
\begin{itemize}
    \item The learning rate scheduler has been changed with respect to the previous work \cite{LopezPortilla:2020odz}, going from a static $lr$ decay, scaled at every training step, to a dynamic one, in which $lr$ is reduced when a metric has stopped improving, $i.e.$ $lr$ changes only if necessary to continue learning.
    \item An early stopping technique has been implemented to allow the network reducing the number of training epochs for easy tasks and, viceversa, increase it for harder ones. In practice, the early stopping procedure consists in monitoring the model's performance on a validation set during training and stopping the training on a given dataset once performance stops improving, $i.e.$ when the validation loss saturates:
    \begin{equation}
        | \pazocal{L}^e_{val} - \pazocal{L}^{e^{best}}_{val} | < \eta, \; \; \text{given} \; \; e-e^{best} \geq n
    \end{equation}
    where $\pazocal{L}^e_{val}$ represents the validation loss of the current epoch $e$, $\pazocal{L}^{e^{best}}_{val}$ is the validation loss of the best epoch $e^{best}$, $\eta$ the tolerance and $n$ the patience. The default values used in this work are $\eta=0.001$ and $n=30$, adapted from \cite{2024arXiv240905068M}.
\end{itemize}
In terms of speed performance, this new software implementation has lead to a significant decrease of the computational time with respect to the previous work. We conduct a test on a GPU NVIDIA A100 using the same amount of data and the same number of epochs (5 epochs for each SNR bin) as in \cite{LopezPortilla:2020odz}, and the model takes around 2 minutes to train, validate and test the network, which has to be compared with the 1h 18min obtained in \cite{LopezPortilla:2020odz} with a GPU NVIDIA P5000.

\subsection{Performance evaluation}
\label{subsec:eff}

Because we are dealing with a binary classification task, the network assigns, to each spectrogram, a scalar value $\theta$ between 0 (noise class) and 1 (signal class), which can be interpreted as the probability for a given spectrogram to contain the injected signal. Commonly, the classification performance of a network is computed from the so-called confusion matrix, whose entries are defined as the number of wrong and right guesses among the test dataset. For a binary classification problem, the confusion matrix is $2\times2$ and consists of the following elements in Table \ref{tab:confusion}.

\begin{table}[h!]
\caption{Confusion matrix of GW + noise (positive class) against only noise (negative class)} \label{tab:confusion}
\begin{tabular}{cc|c|c|c}

\cline{3-4}
&  & \multicolumn{2}{c|}{\textbf{Actual class}}  &  \\ \cline{3-4} & & \textit{GW + noise} & \textit{Noise} &  \\ \cline{1-4}
\multicolumn{1}{|c|}{\multirow{2}{*}{\textbf{\begin{tabular}[c]{@{}c@{}}Predicted\\ class\end{tabular}}}} & \textit{GW + noise} & \begin{tabular}[c]{@{}c@{}}True\\  positive (TP)\end{tabular}  & \begin{tabular}[c]{@{}c@{}}False \\ positive (FP)\end{tabular} &  \\ \cline{2-4}
\multicolumn{1}{|c|}{}                                                                                    & \textit{Noise} & \begin{tabular}[c]{@{}c@{}}False \\ negative (FN)\end{tabular} & \begin{tabular}[c]{@{}c@{}}True\\  negative (TN)\end{tabular}  &  \\ \cline{1-4}
\end{tabular}
\end{table}

These elements are computed by fixing a threshold $\theta^*$ so that images with a value of $\theta > \theta^*$ are classified as signal, while the ones with $\theta < \theta^*$ as noise. From confusion matrix entries, we can define two important quantities: efficiency $\eta$ and false alarm rate $FAR$.
\begin{equation}
    \eta = \frac{TP}{TP+FN}
    \label{eff}
\end{equation}
\begin{equation}
    FAR = \frac{FP}{TN+FP}
    \label{far}
\end{equation}
which express how well we can classify the signal and what are the expected rate of misclassified noise images, respectively. By varying the value of $\theta^*$, we can build the so-called Receiver Operating Characteristic (ROC) curve as the trade-off between $\eta$ and the $FAR$ at various thresholds $\theta^*$.


\begin{figure}
    \centering
    \includegraphics[width=0.7\linewidth]{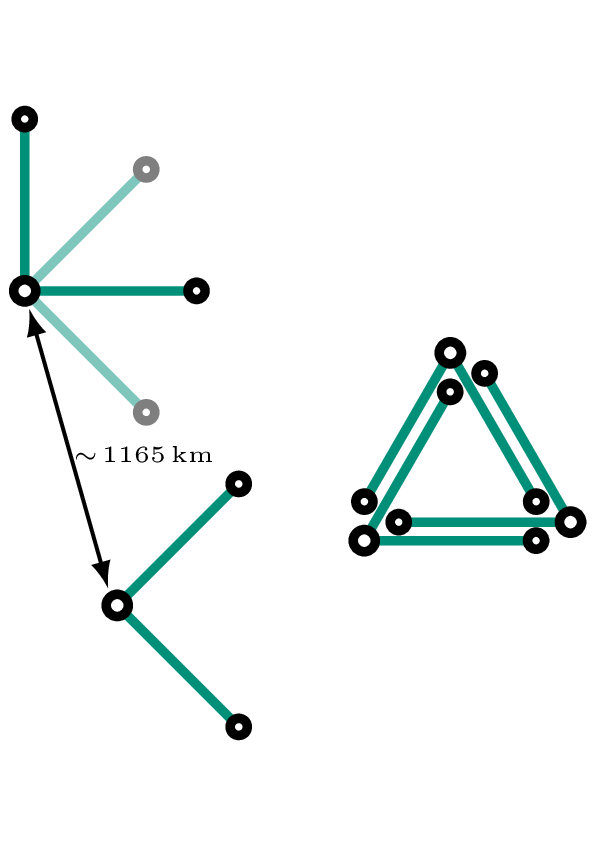}
    \includegraphics[width=\linewidth]{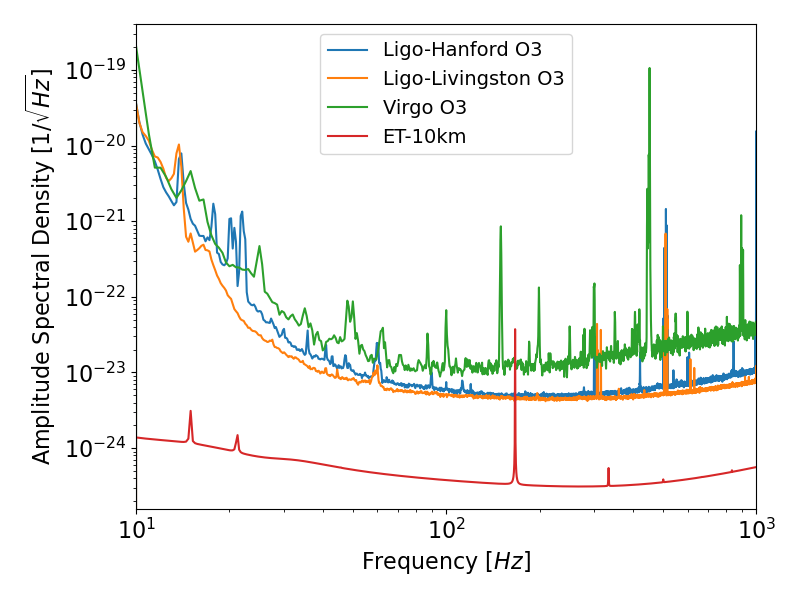}
    \caption{(top) three interferometer configurations among the ones discussed in \cite{Branchesi:2023mws} for Einstein Telescope (ET): triangle configuration and 2L with parallel arms and relative inclination of 45$^\circ$. (bottom) ET sensitivity curve considered in \cite{Branchesi:2023mws} for a 10 km arms interferometer compared with O3 sensitivities.}
    \label{fig:et}
\end{figure}

\section{Dataset}
\label{sec:dataset}

Following the {DL} paradigm, three main datasets are created: training and validation datasets, which are used during the learning phase to tune the model parameters; and test dataset which is applied to compute the network performance after the training. Each of them is composed of positive (signal waveforms injected in detector noise) and negative (pure detector noise) classes, equally distributed in each dataset.

\subsection{Training and Validation}
Training is the process by which a {DL} model learns from data by adjusting its parameters (weights and biases) to minimize the difference between the predicted output and the actual label, while validation helps ensure the model generalizes to unseen data by tuning hyperparameters and preventing overfitting. Data are divided so that 75\% is used for the training and 25\% for the validation. In both cases, signal class data are built using phenomenological waveforms described in Section \ref{sec:phenomenological}.
\subsection{Test}
The test phase is the final step in evaluating a {DL} model, where the network's performance is assessed on completely unseen data that was not used during training or validation. The purpose of the test phase is to provide an unbiased estimate of how well the model is likely to perform in real-world scenarios. In this work, the network test was performed using GW signals from the simulations described in Table~\ref{tab:1} which are selected in base to the next criteria: i) three dimensional simulations, ii) non-rotating progenitors, iii) no-magnetic fields, iv) some neutrino transport treatment, and v) progenitors with single star evolution (we exclude e.g. ultra-stripped stars). This selection is the same as in \cite{Cerda-Duran2025} (see their Table I, for more details). The resulting dataset is composed of 34 different waveforms and, thanks to the wide range of masses and energies spanned, it can be considered a representative sample of what we should expect from a population of real CCSNe.

\section{Results}
\label{sec:results}

\begin{figure}
    \centering
    \includegraphics[width=\linewidth]{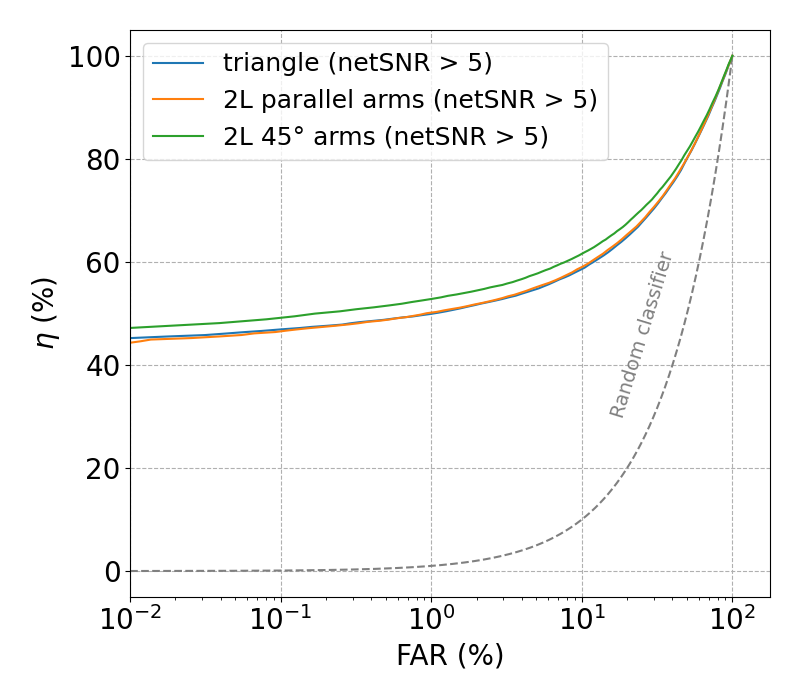}
    \caption{Receiver Operating Characteristic (ROC) curves (efficiency $\eta$ vs against false alarm rate $FAR$) for three interferometer configurations considered in this work. The dashed grey line stands the random classifier, which makes predictions purely by chance, without learning from the data.}
    \label{fig:roc}
\end{figure}

\begin{figure}
    \centering
    \includegraphics[width=0.9\linewidth]{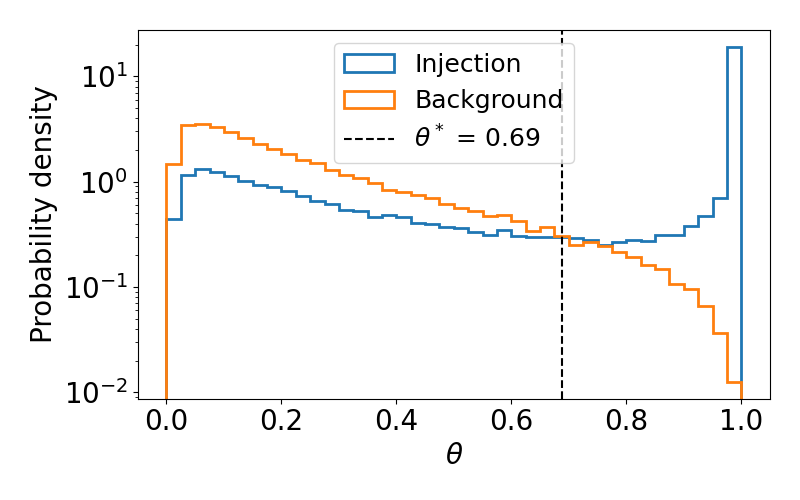}
    \caption{$\theta$ distribution for test dataset of ET 2L with a relative inclination of 45$^\circ$. The vertical dashed line stands for the threshold value that allow to reach a FAR of 5\%. Given the counts of the $i$th bin $c_i$ and its width $b_i$, we define the probability density  as $c_i/(\sum_i^N c_i \times b_i)$, where $N$ is the total number of bins of the histogram.}
    \label{fig:theta}
\end{figure}

Aiming to show the performance of this method, we report the results obtained for third generation ground-based gravitational wave detectors. In particular, we consider the three main detector configurations proposed in \cite{Branchesi:2023mws} for Einstein Telescope (ET): (i) three detectors with a $60^\circ$ opening angle, arranged in the shape of a triangle, (ii) couple of interferometers (2L) with parallel arms (iii) couple of interferometers (2L) with a relative inclination of $45^\circ$. While the three detectors composing the triangle arrangement are located at the same place, the two L interferometers are located at different sites (Sardinia and Netherlands). For each configuration, an ideal Gaussian noise has been generated according to the detector sensitivity reported in \cite{Branchesi:2023mws} for 10 km arms interferometers. In Fig. \ref{fig:et} the design PSD of ET is compared with the sensitivity curves of current GW interferometers achieved during the third observational run (O3). \newline

A total of 50k waveforms have been injected or both the training (plus validation) and test datasets. The source distance have been sampled from a uniform distribution going up to 10 kpc for training and validation, and up to 200 kpc for test.
During the training, different values of network hyper-parameters (learning rate $lr$, batch size $bs$, positive class weight $w_{pos}$) have been explored (see Table \ref{tab:hyperparams} for details). Note that the choice of selecting only values of $w_{pos}$ smaller or equal to 1 is required by the preference of having a model with the lowest FAR, $i.e.$ we want the model to pay more attention to correctly classify the negative class. Moreover, following what was done in \cite{Astone:2018uge} we explored the case in which injections with a very low netSNR are neglected during the training. The reason of this choice is straightforward: because they are completely embedded inside the noise strain even after the whitening, they cannot be distinguished from the pure noise samples. To this extent, three thresholds have been investigated for both the triangle and 2L configuraions: netSNR $>$ [0, 5, 10].

\begin{table}[h]
\caption{Summary of hyperparameter ranges used for fine-tunning.}
\centering
\begin{tabular}{ll}
\hline
\textbf{Hyperparameter} & \textbf{Values} \\
\hline
Learning rate ($lr$)    & $[0.001,\ 0.0001]$ \\
Batch size ($bs$)       & $\{32,\ 64,\ 128\}$ \\
Positive weight ($w_{\text{pos}}$) & $\{0.33,\ 0.50,\ 1.00\}$ \\
\hline
\end{tabular}
\label{tab:hyperparams}
\end{table}

For each detector configuration, among all the possible combinations of hyperparameters and network SNR thresholds, the trained model achieving the highest efficiency is selected. Then, the choice of the best detector configuration is carried on by comparing the ROC curves of the three models, as shown in Fig. \ref{fig:roc}. Note that the fairness of making this kind of comparison is guaranteed by the fact that we are considering similar test datasets, $i.e.$ same distance and sky localization distributions. Among the ones studied in this work, the model achieving the best performance is indeed the 2L with a relative inclination of $45^\circ$ and with netSNR $>$ 5. However, the discrepancy between the curves is so small that would have resulted in a difference of few kpc in detection efficiency. Because the results should not be that different among the three configurations, in the next we will focus our attention only on the 2L with relative inclination of $45^\circ$. \newline

As it has been explained in Section \ref{subsec:eff}, the efficiency of the network $\eta$ can be computed once the value of $\theta^*$ has been established. To do so, a common practice is to set a threshold on the $FAR$ (Eq. \ref{far}) and retrieve the value of $\theta^*$ that allows to reach that threshold. This approach highlights the importance of being aware of the value of $FAR$ when we want to deal with signal detection in astronomy, because it quantifies the probability of spurious detections and thereby constrains the confidence level associated with candidate events.
In this work, the value of $\theta^*$ has been fixed so that a $FAR$ of 5\% is reached (see Fig. \ref{fig:theta}), which results in a value of $\theta^*$\ = 0.69. Note that there is still room to go down in FAR, but this is something that will be explored in future works.

\begin{figure}
    \centering
    \includegraphics[width=\linewidth]{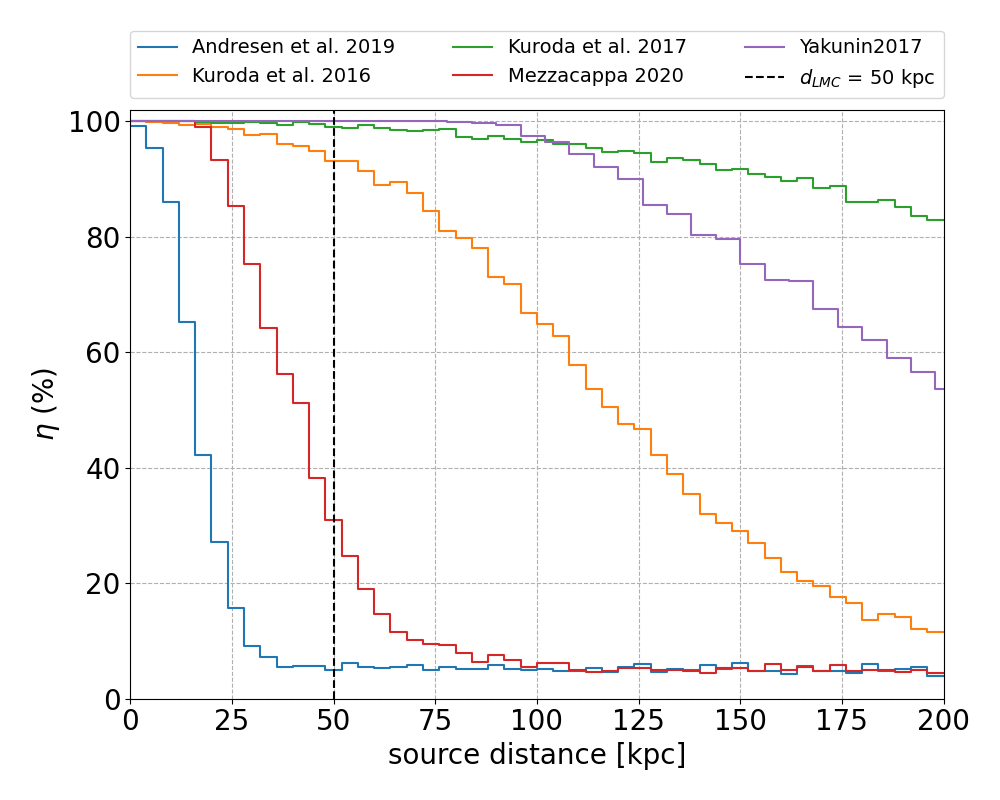}
    \caption{Detection efficiency $\eta$ (Eq. \ref{eff}) in function of source distance for five numerical simulation samples sharing the same progenitor mass of $15M_\odot$: (1) Andresen et al. 2019 \cite{Andresen2019}; (2) Kuroda et al. 2016 \cite{Kuroda:2016}; (3) Kuroda et al. 2017 \cite{Kuroda2017}; (4) Mezzacappa 2020 \cite{Mezzacappa2020}; (5) Yakunin et al. 2017 \cite{Yakunin2017}. The curves are computed considering ET 2L with a relative inclination of $45^\circ$. The dashed line stands for the distance of the Large Magellanic Cloud (LMC).}
    \label{fig:eff}
\end{figure}

\begin{figure}
    \centering
    \includegraphics[width=\linewidth]{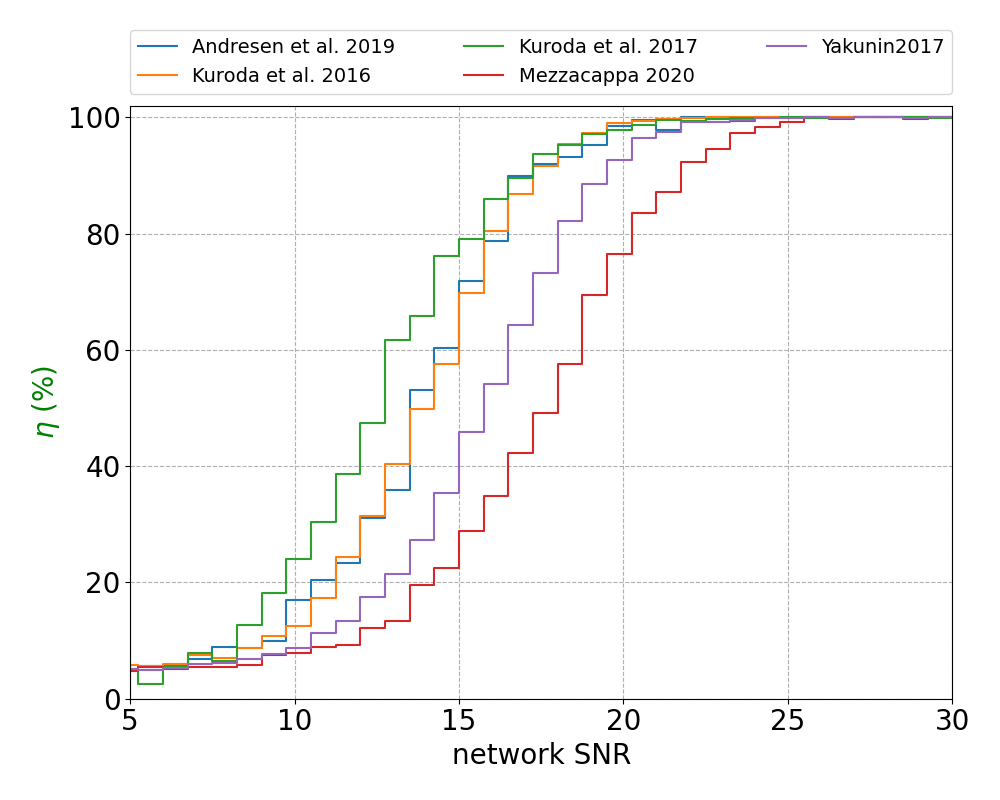}
    \caption{Detection efficiency $\eta$ (Eq. \ref{eff}) in function of network SNR (Eq. \ref{netsnr}) for the selected simulation sample. The curves are computed considering ET 2L with a relative inclination of $45^\circ$.}
    \label{fig:eff_snr}
\end{figure}

For each waveform from the test dataset, we compute the efficiency $\eta$ in function of the source distance up to 200 kpc. With the aim of showing how much the detection capability is affected by the high level of stochasticity of CCSN mechanism, we report in Fig. \ref{fig:eff} the efficiency curves for only five simulation outcomes having the same progenitor mass of $15M_\odot$: Andresen et al. 2019 \cite{Andresen2019}, Kuroda et al. 2016 \cite{Kuroda:2016}, Kuroda et al. 2017 \cite{Kuroda2017}, Mezzacappa 2020 \cite{Mezzacappa2020}, Yakunin et al. 2017 \cite{Yakunin2017}. It is here important to mention that the five simulations were chosen in order to be still representative of what we should expect from a real CCSN. This simply means that the selected dataset still contains the signals achieving the lowest (Andresen et al. 2019) and highest (Kuroda et al. 2017) detection efficiency among the whole test dataset. Table \ref{tab:eff} reports the distance achieved for each waveforms at $\eta=50\%$. Despite sharing the same progenitor mass, the efficiency curves appear very different one to the other. Two are the main reasons behind this behavior. First, there is the loudness of the signal: the louder the signal ($i.e.$ higher root mean square strain $h_{rms} = \sqrt{<h_+^2(t) + h_\times^2(t)>_t}$ where $<\cdot>_t$ denotes the temporal mean), the easier the detection. Indeed, being Kuroda et al 2016 about 5 times louder in $h_{rms}$ than Andresen et al. 2019, results in a higher efficiency over the whole distance interval. However, the loudness is not the only relevant parameter, otherwise Yakunin et al. 2017 would have the highest detection efficiency among the whole test dataset. Instead, above a certain distance, the \textit{complexity} of the signal starts playing a crucial role. Here, "complexity" is used to denote how difficult is to retrieve the main expected features (g-mode and SASI) from the GW strain. This is clear from Fig. \ref{fig:eff_snr} where the detection efficiency is reported as a function of the network SNR for the five waveforms. Although Yakunin et al. 2017 is the loudest one (2.5 louder in $h_{rms}$ than Kuroda et al. 2016), its complexity makes g-mode and SASI hardly distinguishable, thus leading to low efficiency even at high netSNR. On the other hand, for the same value of netSNR, Kuroda et al. 2017, which is "only" 1.8 louder in $h_{rms}$ than Kuroda et al. 2016, achieve the best performance thanks to the clear and unambiguous presence of SASI feature. Finally, it is worth to notice that in both Figure \ref{fig:eff} and \ref{fig:eff_snr}, the detection efficiency at high distances and low netSNR, respectively, saturates at about 5\%, as we should expected from the fact that $\theta^*$ has been chosen in order to achieve a FAR of 5\%. Indeed, at very low netSNR and high distances, the injected signal is indistinguishable from the noise distribution. Therefore, the detection efficiency must converge to the value of FAR previously set.

From the results reported in Fig. \ref{eff}, we can state that, if we assume a signal like the one obtained by Kuroda et al. 2016 and the ET detector configuration composed of a couple of 10 km arm interferometer with an opening angle of $90^\circ$ and relative inclination of 45°, we should be able to reach a detection efficiency above 90\% at 50 kpc, that is where the Large Magellanic Cloud is located. If we consider the best case scenario (Kuroda et al. 2017), the distance at which detection efficiency is around 90\% become $\sim 150$ kpc. This result is competitive with what have been done so far in the context of CCSN search with ET, where the maximum distance of detection for 3G detectors is estimated to be around 100 kpc \cite{Roma:2019kcd, Abac:2025saz, Powell2020}.

\begin{table}[h]
\caption{Summary of ET detection efficiencies. For each waveform, the root mean square strain $h_{rms}$} has been scaled by the one of Kuroda et al. 2016.
\centering
\begin{tabular}{lll}
\hline
\textbf{Waveform} & $h_{rms}$ & \textbf{Distance at $\eta = 50\%$} \\
\hline
Andresen 2019    & 0.19 & 17 kpc \\
Kuroda 2016      & 1.0 & 120 kpc \\
Kuroda 2017      & 1.82 & $> 200$ kpc \\
Mezzacappa 2020  & 0.85 & 42 kpc \\
Yakunin 2017     & 2.50 & $\approx$ 200 kpc \\
\hline
\end{tabular}
\label{tab:eff}
\end{table}

\section{Conclusion}
\label{sec:conclusion}

Multi-messenger Understanding of Supernova Explosions (MUSE) is a pipeline developed with the aim of searching for CCSN signatures in GW data. It is based on a Convolutional Neural Network trained on phenomenological waveforms, that are built to mimic the main GW emission features expected from numerical simulation outcomes. The final version of the model is an upgrade of what is described in \cite{LopezPortilla:2020odz}. In this work, MUSE performances have been probed on Einstein Telescope, a third generation GW detector. Three configurations have been studied: (i) three detectors with a $60^\circ$ opening angle, arranged in the shape of a triangle, (ii) couple of interferometers (2L) with parallel arms (iii) couple of interferometers (2L) with a relative inclination of $45^\circ$. 
The predicted detector sensitivity for a 10 km arms interferometer has been adopted.
This study has shown that the detector configuration with the best performance is the 2L with a relative inclination of $45^\circ$. This configuration is found to have the best performance also in the context of CBC searches \cite{Branchesi:2023mws}.
For that specific configuration, we computed the detectability of ET for a representative set of 3D CCSN simulations
and we showed that, with MUSE, we would be able to detect a CCSN with the same property as the one of Kuroda et al. 2016 (SFHx), that is hosted in the Large Magellanic Cloud (at 50 kpc from us), with an efficiency above 90\%.  \newline
In future works, we plan to use MUSE on real data starting with the ones from the third observational run (O3) of the LIGO and Virgo detectors. The test sample will also be extended, taking into account more recent numerical simulation than the one presented in this work. The phenomenological template will also be upgraded with the inclusion of magneto-rotational, which represent only about 1\% of all CCSNe but are expected to produce louder GW emission and so easier to detect at even higher distances.

\section{Acknowledgements}
\label{s:acknow}

IDP acknowledges the support of the  Sapienza  School for Advanced Studies (SSAS) and the support of the Sapienza Grant No.RM124190B27960E5. MD acknowledges the support of the Sapienza Grants No. RM123188F3F2172A and RM1221816813FFA3. PCD acknowledges the support from the grants 
Prometeo CIPROM/2022/49 from the Generalitat Valenciana, and  PID2021-125485NB-C21 from the Spanish Agencia Estatal de Investigación funded by MCIN/AEI/10.13039/501100011033 and ERDF A way of making Europe. ML is supported by the research program of the Netherlands Organization for Scientific Research (NWO). The authors are grateful for computational resources provided by the LIGO Laboratory and supported by the National Science Foundation Grants No. PHY-0757058 and No. PHY-0823459. 

\bibliography{references}
\end{document}